\documentclass[preprint,authoryear,12pt]{elsarticle}

\usepackage{epsfig}

\usepackage{amssymb}

\usepackage[ps2pdf,%
a4paper=true,%
breaklinks=true,%
colorlinks=true,%
pdfauthor={First Author et al.},%
pdftitle={Template for manuscripts in Advances in Space Research}%
]{hyperref}

\begin{document}

\begin{frontmatter}



\title{Measurement of secondary cosmic ray intensity at Regener-Pfotzer height using low-cost weather balloons and its correlation with solar activity}

\author[icsp]{Ritabrata Sarkar\corref{cor}}
\ead{ritabrata.s@gmail.com}
\cortext[cor]{Corresponding author}

\author[snb,icsp]{Sandip K. Chakrabarti}
\ead{sandipchakrabarti9@gmail.com}

\author[icsp]{Partha Sarathi Pal}
\ead{parthasarathi.pal@gmail.com}

\author[icsp]{Debashis Bhowmick}
\ead{debashisbhowmick@gmail.com}

\author[icsp]{Arnab Bhattacharya}
\ead{arnabseacom@yahoo.com}

\address[icsp]{Indian Centre for Space Physics, 43 Chalantika, 
           Garia Station Rd., Kolkata 700084, West Bengal, India}

\address[snb]{S.N. Bose National Centre for Basic Sciences,
           JD Block, Salt Lake, Kolkata 700097, West Bengal, India}

\begin{abstract}
Cosmic ray flux in our planetary system is primarily modulated by solar
activity. Radiation effects of cosmic rays on the Earth strongly depend
on latitude due to the variation of the geomagnetic field strength. To
study these effects we carried out a series of measurements of the radiation
characteristics in the atmosphere due to cosmic rays from various places
(geomagnetic latitude: $\sim 14.50^{\circ}$ N) in West Bengal, India, located near
the Tropic of Cancer, for several years (2012-2016) particularly covering
the solar maximum in the 24th solar cycle. We present low energy ($15-140$
keV) secondary radiation measurement results extending from the ground till
the near space ($\sim 40$ km) using a scintillator detector on board rubber
weather balloons. We also concentrate on the cosmic ray intensity at the
Regener-Pfotzer maxima and find a strong anti-correlation between this intensity
and the solar activity even at low geomagnetic latitudes.
\end{abstract}

\begin{keyword}
Cosmic ray \sep Regener-Pfotzer height \sep Solar activity \sep X-ray detector \sep
Weather balloon-borne mission
\PACS 94.20.wq \sep 94.05.Sd \sep 95.55.-n \sep 95.55.Ka \sep 96.60.Q
\end{keyword}

\end{frontmatter}

\parindent=0.5 cm

\section{Introduction}
\label{intro}
Interaction of primary Cosmic Rays (CR) with the atmospheric gas nuclei
produces secondary particles and radiations which can be measured in situ
by radiation detectors giving rise to an indirect measurement of the primary.
This is the dominating radiation in the upper atmosphere below an altitude
of about $60$ km. At heights close to the ground (say, below $\sim 2$ km),
the radiation field is dominated by terrestrial radioactive sources \citep{bazi00}.

The secondary CR cascade starting at the upper atmosphere gradually gets
intensified at lower heights and have a maximum radiation at a height about
$15-20$ km depending on the latitude \citep{bazi00}, where the generation
of the secondary radiation is balanced by the loss effects. This effect was 
observed by direct measurement of ionization at various heights in the
atmosphere almost a century ago by Erich Regener and George Pfotzer
\citep{pfot36, rege33, rege35}. This point of maximum radiation is now
called the Regener-Pfotzer maximum (RP-max) and below this point the secondary
radiation intensity is gradually reduced due to absorption and decay processes
\citep{gais90, grie01}.

The intensity of the primary CR, on the other hand, depends on the strength
and spatial distribution of the Earth's magnetic field \citep{stor55}. This
is a highly dynamical system which is influenced by the solar wind and the
interaction between the terrestrial and Interplanetary Magnetic Fields (IMF).
Solar Energetic Particles (SEP) generated in the solar atmosphere due to
various explosive processes can significantly affect the space environment
and produce hazardous radiation effects disturbing the satellite operations,
manned or robotic missions in the low-earth orbits often  affecting the
crews and passengers in commercial aircrafts \citep{icsp, miro03}.

Strong perturbations of the geomagnetic field due to the earthward directed
SEPs can culminate in geomagnetic storms due to large transfer of solar wind
energy to geomagnetic fields. This can significantly change the current,
plasma and the field intensity as well as its structure. The intense
geomagnetic storms can affect geomagnetic shielding and thus affect the
distribution of the planetary CR \citep{dorm71}. Due to this shielding
effect, the location and the amount of radiation at the RP-max and in 
the stratosphere were found to depend on the geomagnetic latitude 
\citep{bazi98, gole77, gole90, shea87}. This secondary radiation also varies 
with solar activity \citep{char75, char79} but this variation is more prominent at 
higher latitudes and has minimal effect at the low-latitude locations where 
the geomagnetic rigidity exceeds $10$ GV. Results reported in the present 
paper are performed in this low-latitude region ($\sim 14.5^{\circ}$ N).

In situ measurements of the atmospheric parameters through various heights 
using balloons and aircrafts is a very old subject of interest. Many attempts 
have been made to measure electrical measurements in the atmosphere
\citep{nico12} and to measure radiation levels at various
altitudes using radiosonde, mainly at mid and high-latitudes
\citep[e.g.,][in Finland]{hata00} \citep[][in Russia]{bazi98} 
\citep[][in UK]{harr14} \citep[][in Israel]{yani16}. However,  the
vertical profile of the atmospheric radioactivity data at low geomagnetic
latitude around the Tropic of Cancer is not available except in one or two
stray cases \cite[e.g.,][measured in Hong Kong sky]{li07}. Thus our
measurement is expected to extend the study to a wider range of latitudes.

\section{Mission description}
\label{sec:miss}
In order to measure secondary radiation due to the CR interaction with
atmospheric matter, Indian Centre for Space Physics (ICSP), Kolkata, India
has been launching a `Dignity' series of Missions on a regular basis.
In these experiments, very light weight payloads (under $2-3$ kg category) are
launched into the atmosphere on board rubber weather balloons which
can attain a height of about 40 km. The main instrument used in these
missions is a scintillator detector for the detection of X-rays. The details
of the detector is given in \S \ref{sec:det}. The payload consists of several
other ancillary equipments, such as: Global Positioning System (GPS) unit for
the location and altitude measurement; payload Attitude and Heading Reference
System (AHRS) using micro electro-mechanical Inertia Measurement Units (IMU)
to measure the payload attitude; other sensors for measuring payload's
internal and external temperature and atmospheric pressure; optical video
camera unit for the recording of overall mission performance. A general
overview of these missions can be found in more detail in \citet{chak14, chak17}.

The main detector unit along with its electronic readout system, data
storage system, power supply and other ancillary equipments are housed
inside a polystyrene (thermocol) box to give the instruments a
shielding from extremely low temperature environment during the entire
flight path and also to protect the components from mechanical shock during
the landing of the payloads. The payload box is generally cylindrical shaped
with overall height of $\sim 36$ cm and diameter $46$ cm. The cylindrical
shape of the payload box was adopted in order to reduce the drag effects.

While ICSP is engaged in various near space experiments specifically
measuring X-rays from compact objects, several missions were
dedicated to measure cosmic ray intensity and spectrum for several years to
collect the data on the nature of secondary CR and to study its seasonal or
yearly variations under diverse solar conditions. Depending on the
tropospheric and stratospheric wind condition, which defines the course of
the payload, we launch the payloads from specific places during specific
times in the year. In Table \ref{tab:mission} the details of mission launch
timings and positions concerning this current work are given.

\begin{table}[h]
  \caption{Mission timing and positions reported in this paper. The actual mission number in the
  Dignity series is mentioned within the parenthesis along with the mission
  index used in this paper.}
  \centering
  \scriptsize
  \begin{tabular}{l c c c c}
  \noalign{\smallskip}
  \hline
  {\bf Mission} & {\bf Date} & {\bf Launch} & {\bf Landing} & {\bf Launch
  location\textsuperscript{$a$}} \\
  {\bf index.} &  & {\bf time (UT)} & {\bf time (UT)} & {\bf (Lat,
  Lon)} \\
  \hline
  \noalign{\smallskip}
  1(D29) & 04-06-2012 & 03:30:19 & 07:11:26 & Bhaluka (23.35N, 88.40E) \\
  2(D45) & 24-10-2013 & 05:46:25 & 08:35:50 & Suri (23.91N, 87.52E) \\
  3(D48) & 06-11-2013 & 04:09:57 & 07:01:57 & Kulti (23.73N, 86.84E) \\
  4(D50) & 17-11-2013 & 04:28:30 & 07:44:47 & Kulti (23.73N, 86.84E) \\
  5(D52) & 17-11-2013 & 19:49:09 & 01:50:27\textsuperscript{$b$} & Kulti
  (23.73N, 86.84E) \\
  6(D57) & 17-05-2014 & 03:48:39 & 08:26:50 & Bolpur (23.67N, 87.69E) \\
  7(D59) & 19-05-2014 & 05:32:47 & 14:08:00 & Bolpur (23.67N, 87.69E) \\
  8(D79) & 17-05-2015 & 04:49:09 & 07:41:00 & Muluk (23.65N, 87.71E) \\
  9(D87) & 21-11-2015 & 04:52:33 & 07:52:04 & Muluk (23.65N, 87.71E) \\
  10(D90) & 09-05-2016 & 04:08:24 & 08:23:49 & Muluk (23.65N, 87.71E) \\
  11(D94) & 13-05-2016 & 04:24:22 & 08:51:24 & Muluk (23.65N, 87.71E) \\
  12(D98) & 17-10-2016 & 03:53:07 & 06:46:01 & Muluk (23.65N, 87.71E) \\
  \noalign{\smallskip}
  \hline
  \noalign{\smallskip}
  \multicolumn{5}{l}{\textsuperscript{$a$}\footnotesize{W.B., India.}} \\
  \multicolumn{5}{l}{\textsuperscript{$b$}\footnotesize{Next day.}} \\
  \end{tabular}
  \label{tab:mission}
\end{table}

\section{Detector description}
\label{sec:det}
\begin{figure}[h]
  \centering
  \includegraphics[width=0.7\textwidth]{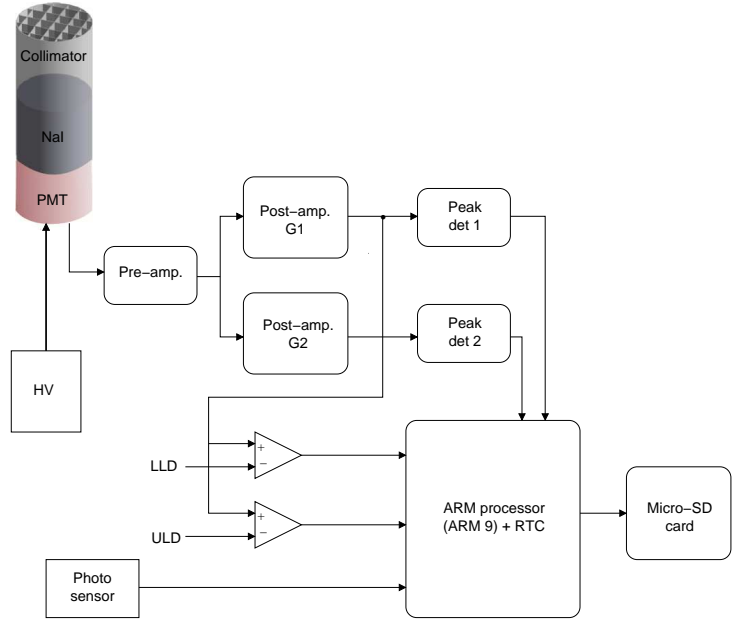}
  \caption{Schematic diagram of the scintillator detector and the event
  readout system comprising 2''$\times$2'' NaI crystal fitted with a
  $10^{\circ}$  FoV lead collimator.}
   \label{fig:det}
\end{figure}

The scintillator detector used to detect the secondary CR radiation is made
of a single NaI(Tl) cylindrical crystal of 2'' diameter and 2'' thick. This
crystal is mounted on a compatible Photo-Multiplier Tube (PMT) of similar
diameter and hermetically sealed in an aluminum housing. The whole integrated
detector assembly (crystal~+~PMT) (model name Bicron Monoline 2M2/2, hereafter
B2 detector) is a product of Saint-Gobain Crystals \citep{saint}. The
front-end electronics system concerning the analog pulse amplification,
shaping etc. and the digital electronics part for the digital pulse processing 
and event storing were designed and fabricated in our laboratory according to
the mission requirements. A schematic drawing of the detector fitted with the
collimator and the readout system is shown in Figure \ref{fig:det}. Though a
collimator is not needed in a CR measurement, we kept it for studies of
energetic sources. So CR measurements are bi-products of
such measurements.

The lower limit of the operating energy range of the detector is defined by
the thickness of the detector housing along with the PMT bias voltage and
the threshold in the discriminator. The window of the detector is made of
$0.5$ mm thick aluminum which permits about $30$\% X-ray at $15$ keV. Also
the extra-terrestrial X-rays below this limit get absorbed in the residual
atmosphere corresponding to the maximum attainable height. The upper limit
of the energy range is about $\sim 5$ MeV as recommended by the detector
manufacturer. But due to our science objectives we are more concerned of the
lower energy data requiring better energy resolution in this region. To
fulfill this purpose we used two post-amplifiers with different gain settings
in the analogue pulse processing part: one (G1) dealing the events in the
energy range about $15-140$ keV and the other (G2) working in the range about
$0.1-2$ MeV. The pulses from the two amplifiers are shaped and converted to
digital signal to be stored in a memory card. Two different trigger circuits
are dedicated for the trigger generation in two energy ranges.

Collimators made by $0.5$ mm thick lead sheets are used to reduce background
radiation counts. This is particularly useful for observation of specific
sources which we observe along the way. The same lead sheet is also used
to shield the detector from radiation entering from sides. Since the lead
sheet is capable of blocking $\sim 100$ keV radiations, above $100$ keV, the
collimators do not function. Thus we concentrate on the radiations below
$60-70$ keV for our purpose. In case any mission has a different collimator
configuration, we normalize all of them using a reference detector.

\section{Data analysis}
\label{sec:ana}
The detectors are provided by collimators of $0.5$ mm thick lead with various
field of view specific to the Mission. The cylindrical side of the detector is
shielded by lead sheet of same thickness. This lead shielding essentially
can provide $\sim 95$\% blocking of photons up to $100$ keV but due to the
absorption line at $88$ keV and K$_{\alpha}$ emission at $75$ keV, this
collimator effectively blocks up to $60$ keV. This defines the upper limit
for the energy consideration of our analysis. We restrict the lower energy
limit to $25$ keV to avoid the noise from the detector at lower energies.

The shielding cannot block high energy photons and these photons pass through
the detector crystal to deposit their energy producing new counts. Since the
high energy photons are more likely to deposit partial energy (in Compton
scattering or pair production) and we do not have any veto detector to discriminate
these partial depositions, the detector counts could be contaminated by such
photons. So, we do not use photon counts in this range in determining the
temporal variation.

\subsection{Count normalization}
\label{ssec:norm}
Since the same detector has been used for several years, we have been
particularly careful to ensure that the variation of the efficiency of the
detector over the years due to bias voltage change or gain shift are
corrected for. Also different configurations of the collimator were used
during different missions. Due to these facts, the detector radiation counts
need to be normalized with appropriate solid angle.

The radiation counts ($C_d$) obtained by the detector 
have two components in the energy range under consideration: (i) Number of photons entered into the
detector through the FoV ($C_f$) of the collimator and deposited its full
energy. This depends on the FoV of the collimator. (ii) Number of high
energy photons ($C_p$) entered into the detector through all directions and
deposited partial energy into it. This depends on the matter distribution
of the detector construction. So we have,

\begin{equation}
C_d = C_f + C_p.
\end{equation}

To separate full and partial energy deposition counts we take the help
of a reference detector which is capable of separating only the number of
full energy depositing photons ($C_{d,r}^m$; here d - used for detected
photons, r - for reference detector and m - for an altitude of minimum
count explained in the following paragraph). This
reference detector is a phoswich detector using $3$ mm thick NaI and $25$ mm
thick CsI scintillator crystals of $11.6$ cm diameter in combination with
a single PMT for signal readout. Thus by operating both the detectors (B2
and the reference detector) under the same conditions number of photons can
be calculated with full energy deposition in the B2 detector from the
detector characteristics (area, quantum efficiency, FoV) of both the
detectors.

To normalize the counts across the missions, we make an assumption that the minimum photon
counts due to CR at about $1$ km above the ground remains more or less the same
for all the missions. This is quite evident in the plot shown in Figure
\ref{fig:gnd}. Thus the full energy deposition counts in the B2 detector at
the Height of Minimum CR (HMCR) may be calculated as,
\begin{equation}
C_f^m = C_{d,r}^m \frac{A_f}{A_r} \frac{Q}{Q_r} \frac{\Omega}{\Omega_r} 
\end{equation}
and corresponding partial energy deposition counts are,
\begin{equation}
C_p^m = C_d^m - C_f^m,
\end{equation}
where, $C_d^m$ is the overall counts detected at HMCR in the mission, $Q$
and $Q_r$ are the quantum efficiency of B2 and reference detector in the
specified energy, $\Omega$ and $\Omega_r$ are the solid angle subtended by
the FoV of the collimator of the two detectors, $A_f$ and $A_r$ are the
surface area of the detectors window with collimator opening. $C_{d,r}^m$
can be obtained from a mission with the phoswich detector on board the payload.

From the ratio of the partial and full energy deposition counts at HMCR we
can calculate the full $C_f^p$ and partial $C_p^p$ energy deposition counts
at the RP-height.
\begin{equation}
C_f^p = \frac{C_f^m C_d^p}{C_d^m},
\end{equation}

\begin{equation}
C_p^p = C_d^p - C_f^p. 
\end{equation}

The secondary cosmic gamma-ray in the atmosphere depends on the zenith angle
or tilt angle of the detector $\theta_t$ \citep{shon77}. Moreover, this
$\theta_t$ dependence again is a function of the altitude as shown in
\citet{bazi98}. So an index of angle dependence $h$ is needed in the
calculation to incorporate this function. Using the above values, the
normalized photon counts at the RP-max can be calculated as,
\begin{equation}
C_n^p = \frac{C_f^p}{Q A_f \Omega (\cos\theta_t)^h} + \frac{C_p^p}{Q A_a 4\pi}
\label{eqn:norm}
\end{equation}
where, $A_a$ is the overall surface area of the crystal in B2 since high
energy photons can enter into the detector from the whole $4\pi$ solid angle.
The effect of the zenith angle from the vertical axis has been taken care 
by the $(\cos\theta_t)^h$ term. The unit of this normalized counts is
$photons/s/cm^2/sr$. However, for practical purpose, in the subsequent
calculation we have considered the omni-directional counts in unit of
$photons/s/cm^2$.

\subsection{Height of Regener-Pfotzer maximum}
\label{ssec:pfot}
The Regener-Pfotzer height (RP-height) at the region close to the launching site is
estimated by first extracting the average count rate over $100$ m height,
then a running average of these count rates is taken at different heights
to smooth the curve. Next a cubic spline is fitted to determine the position
of the maximum count for the RP-max height and corresponding count rate as the Regener-Pfotzer count. It is to be noted that our flights laterally move by a distance of
about $20-30$ km by the time RP-height is reached. So there is negligible
influence of variation of latitude between the launching and RP-height coordinates.

\section{Results and discussions}
\label{sec:res}

\subsection{Geomagnetic latitude effect}

\begin{figure}[h]
  \centering
  \includegraphics[width=0.7\textwidth]{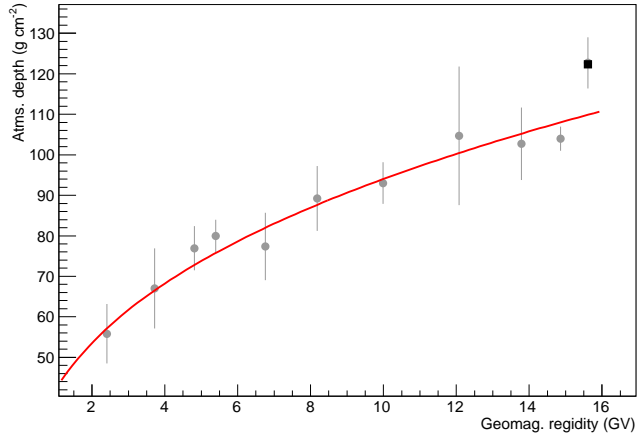}
  \caption{Atmospheric depth at Regener-Pfotzer height vs. geomagnetic rigidity at
  various locations. The result of the location of current study using a
  scintillator detector in a narrow energy band is shown with the black
  square. Other results use Geiger-M$\ddot{\rm u}$ller counters of
  unrestricted energy band and the curve corresponds to an empirical
  fit to these results.}
  \label{fig:rig}
\end{figure}

All the experiments under consideration of this work were conducted near the
Tropic of Cancer as can be seen from Table \ref{tab:mission}. It is well
known that there is a strong dependence of the RP-height with geomagnetic
latitude due the variation of cut-off rigidity of the primary cosmic rays
entering into the atmosphere \citep{bazi98}. The average RP-height found in
this work is $15.0\, \pm\, 0.32$ km. The corresponding atmospheric depth is
calculated as $122.70\, \pm\, 6.33$ g/cm$^{2}$ considering the atmospheric depth
and altitude relation for standard atmosphere as given in \citet{bazi00}.
The geomagnetic rigidity cut-off at this latitude is $15.6$ GV, calculated
according to \citet{bobi06} using the web based calculator \citep{geomag}.
The dependence of the RP-height in terms of atmospheric depth $p$ with the
geomagnetic rigidity cut-off $R_c$ is  shown in Figure \ref{fig:rig}.
Our data in a restricted energy range is represented by a filled square and
the data points obtained from Geiger-M$\ddot{\rm u}$ller counters for other
latitudes are taken from \citet{bazi98} and \citet{li07}. \citet{li07}
used the following empirical formula to fit data:

\begin{equation}
p = 42\, R_c^{0.35},
\end{equation}
shown by the curve in Figure \ref{fig:rig}.

Apart from the difference in energy ranges in consideration, unlike these
results, our result is by far in the region closest to the geomagnetic
equator where the behaviour could be different. Because of these, we
believe that out result is not actually falling on the curve proposed by
\citet{li07}.

\subsection{Ground data comparison}
\begin{figure}[h]
  \centering
  \includegraphics[width=0.7\textwidth]{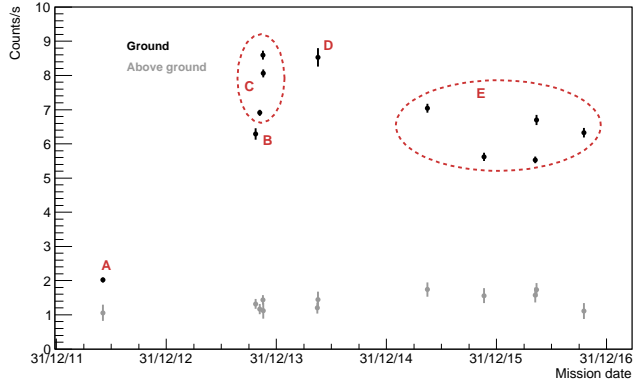}
  \caption{Ground counts (black) and minimum counts above ground (gray) in
  the missions at various places grouped and marked. A - Bhaluka, B - Suri,
  C - Kulti, D - Bolpur, E - Muluk (see Table \ref{tab:mission} for details)
.}
  \label{fig:gnd}
\end{figure}

Radiation counts at the ground or sea level mainly depends on the local 
abundance of radioactive materials and radioactivity of the aerosols at 
that place \citep{ages81}. Radiation counts at the lower atmosphere
also depends on the seasonal variation and local weather effects like rain,
temperature and its distribution etc. In Figure \ref{fig:gnd} we plot the
ground radiation counts for different missions where the data points for
the same locations are grouped in the Figure. The ground radiation and local
effects gradually vanish with altitude in $1-2$ km and secondary radiation
due to CR starts to dominate. So there is a local minimum in the radiation
count profile at that height. The count at the minimum is more or less
the same as evident from the same Figure.

\subsection{Vertical profile of secondary radioactivity}
\begin{figure}[h]
\centering
\includegraphics[width=0.50\textwidth]{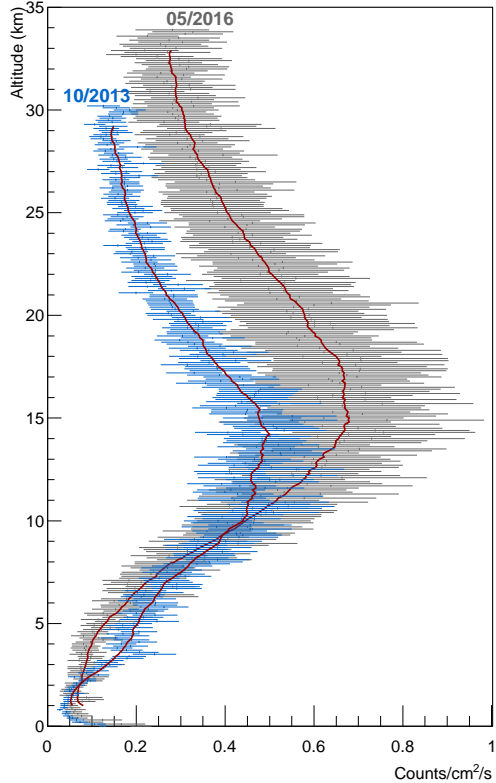}
\caption{Vertical profile of the secondary cosmic ray counts. Two different
data sets from two missions during different phases of solar activity are
shown: (blue) October, 2013 when the Sun undergoing maximum activity and
(gray) May, 2016 after the solar maxima period.}
\label{fig:verprof}
\end{figure}

The vertical profile of radioactivity in atmosphere is shown in Figure
\ref{fig:verprof}, where the normalized radiation count rates per detector
area are plotted in $100$ m intervals along with the running average
presented by the thick lines. In general, the ground radioactivity and the
total count decrease sharply up to about $1$ km due to absorption of the
ground radiation. The total count starts to increase up to the RP-height due
to the increase in secondary radiation for CR interaction in atmosphere and
decrease in absorption. Beyond the RP-max the counts decrease due to reduced
interaction probability of the CR particles in thinner atmosphere.

To demonstrate variation in the atmospheric radiation profile for different
solar conditions, we considered two data sets from two different missions
during different phases of solar activity. The atmospheric radioactivity at
higher altitudes distinctly shows lower counts for the mission during the
solar maxima (October, 2013; Serial No. 2) compared to the data during
(May, 2016; Serial No. 11) when the solar maximum is over. 
Similar type of variation in atmospheric radioactivity during
different phases of solar activity also has been reported from measurments
over Israel \citep{yani16}, Murmansk (Russia) and Antarctica \citep{char75}.
The result is clearly due to the direct consequence of the 
regulation of the primary cosmic rays by solar winds.

\subsection{Variation of Regener-Pfotzer height}
\begin{figure}[h]
  \centering
  \includegraphics[width=0.7\textwidth]{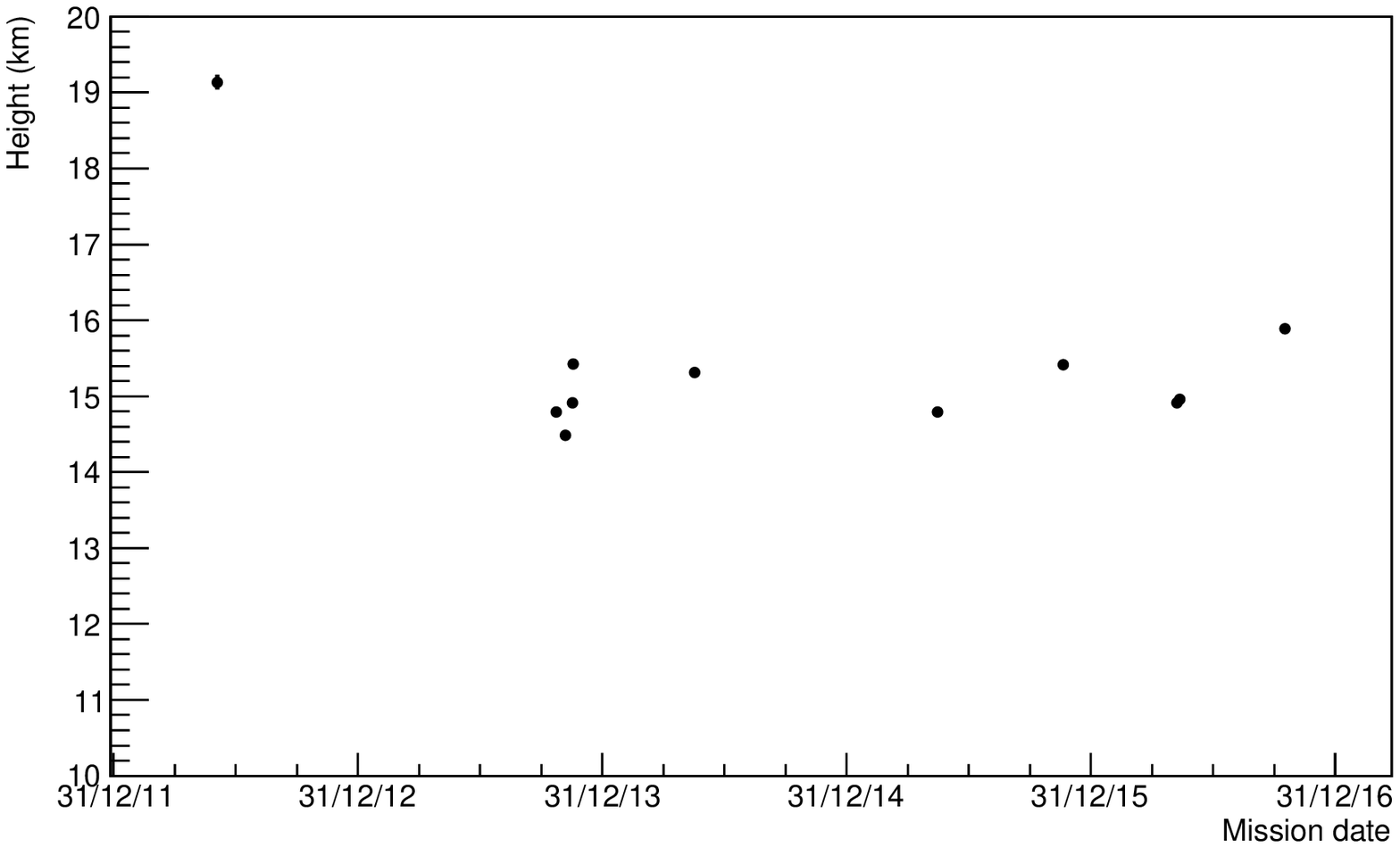}
  \caption{Regener-Pfotzer height obtained for various missions from the secondary
  cosmic-ray counts variation with altitude.}
  \label{fig:phgt}
\end{figure}

Figure \ref{fig:phgt} shows the RP-heights obtained in different missions.
Except one point from the mission (no. 1) in June, 2012 at latitude
$23.35^{\circ}$ N all the other RP-heights have their mean at $15.0\, \pm\, 0.32$
km while the missions took place in the geographic latitude range of
$23.65-23.91^{\circ}$ N. This corresponds to the variation of geomagnetic
latitude of $14.44-14.70^{\circ}$ N. It is possible that due to excess of
solar activity in 2012 during solar maximum, the inseparable solar radiation
also contributed to CR.

\subsection{Effects of solar activity on secondary cosmic rays}

\begin{figure}[h]
  \centering
  \includegraphics[width=0.60\textwidth]{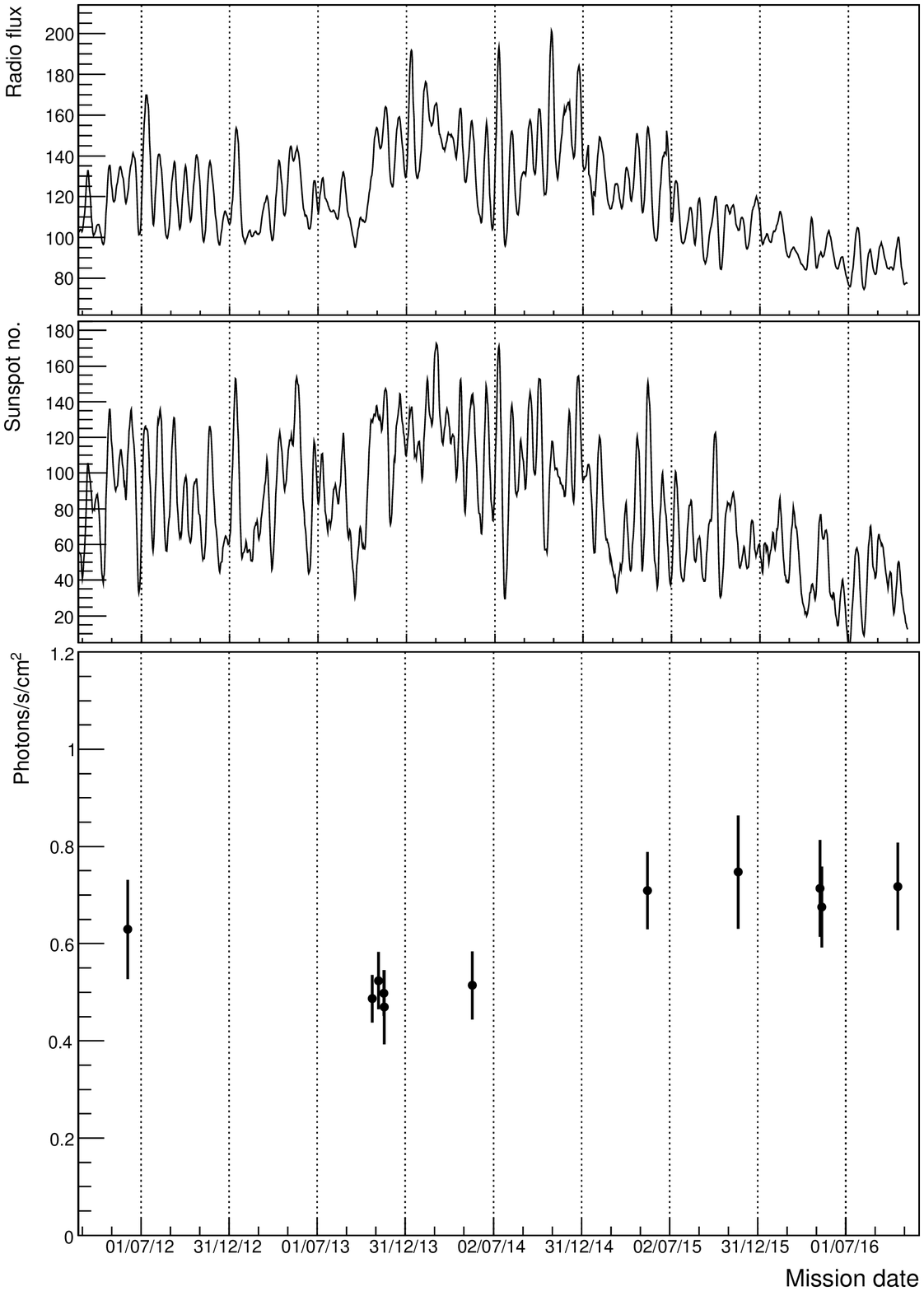}
  \caption{Variation of solar $10.7$ cm radio flux (in unit of
  $10^{-22}\, Watt\, m^{-2} Hz^{-1}$) and the sunspot numbers (both averaged
  over previous $10$ days) along with CR secondary photon counts at Regener-Pfotzer
  maximum in different missions.}
   \label{fig:spcnt}
\end{figure}

To study the effects of solar activity on the atmospheric radiation due to
CRs, the radiation count rates at the RP-max in different missions are plotted
in Figure \ref{fig:spcnt}, along with the sunspot numbers and the $10.7$ cm
radio flux both with a moving average over previous $10$ days. As evident
from the Figure the solar parameters show an anti-correlation with the radiation 
counts at RP-max.

\begin{figure}[h]
  \centering
  \includegraphics[width=0.7\textwidth]{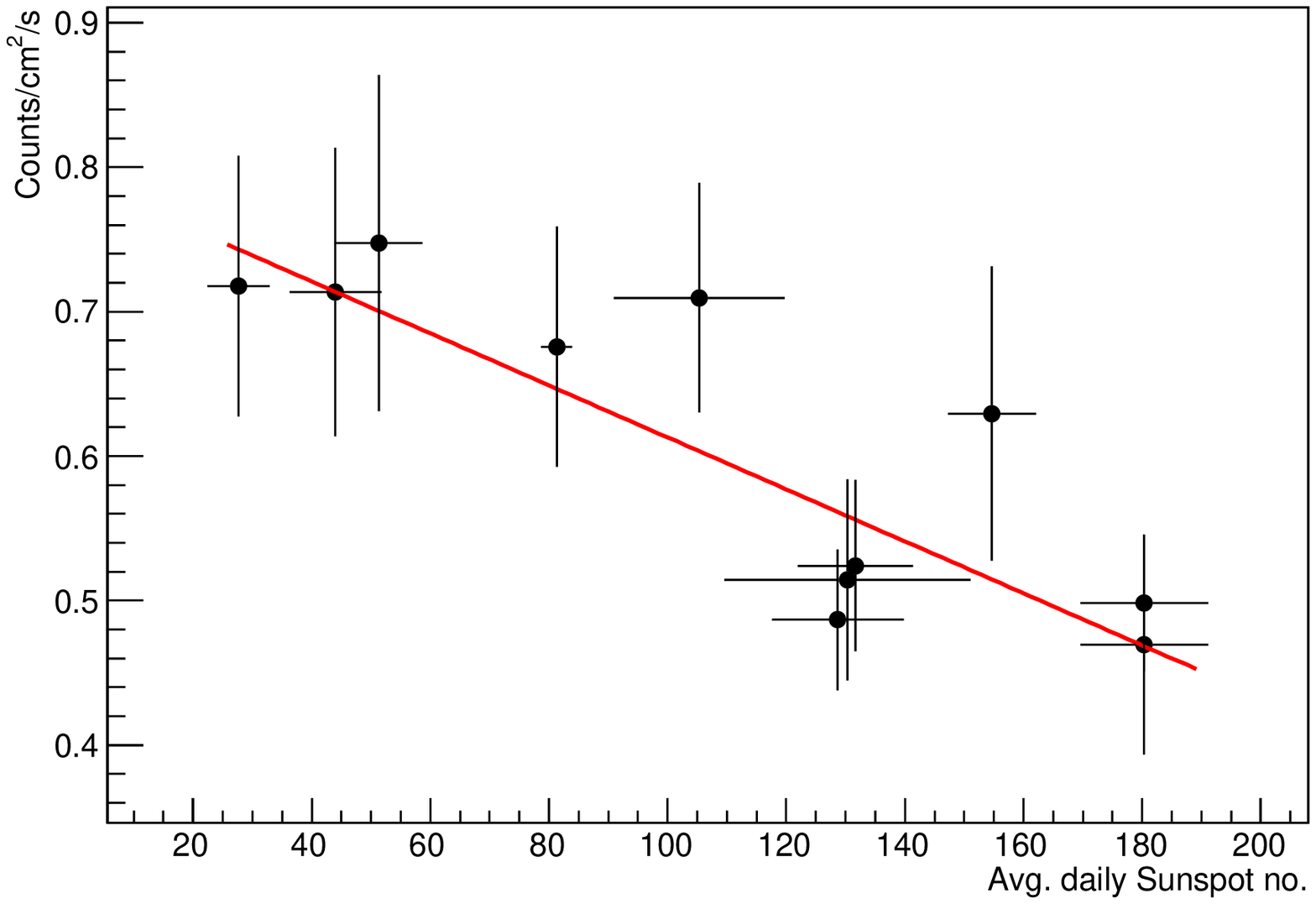}
  \caption{Correlation between the daily sunspot number averaged over previous
  $3$ days and radiation counts at the Regener-Pfotzer maximum.}
   \label{fig:corncss}
\end{figure}

The correlation between daily sunspot number ($S$) averaged over previous
$3$ days and radiation count rate ($C$) at the RP-max is shown in Figure
\ref{fig:corncss}. The plot is fitted with the straight line:
\begin{equation}
C = 0.793 - 0.0018\, S
\end{equation}
with a reduced $\chi^2$ value of $6.14/9$. This shows a strong
anti-correlation with a correlation coefficient of $-0.84$ and a t-test
indicates the anti-correlation is statistically significant at the $95$\%
confidence level.

\section{Conclusions}
\label{sec:conc}
A series of unique low cost experiments using light weight scintillator
detector on board rubber weather balloons have been conducted to measure the
solar effect on the secondary cosmic rays for a time period covering the
solar maximum of the 24th solar cycle. Location of these experiments were
confined near the Tropic of Cancer to measure these effects, even at
relatively low geomagnetic latitude where the rigidity cut-off for the
primary cosmic rays is very high. The dependence of the RP-height with the
rigidity cut-off match generally well with the trend obtained for the same
at other higher latitudes \citep{bazi98, li07}. Results from
the missions at different solar phases, however shows that the RP-height itself
does not change significantly with the solar activity. We noticed a
significant effect of solar activity on the radiation counts at RP-max. We
find that the count rate is anti-correlated with the solar parameters
directly related to the solar activity. \citet{li07} also reported a slight
decrease of the count rate with increasing sunspot number over Hong Kong
sky. \citep{yani16} found some similar results as reported here from 
the measurment of vertical ionization profile over southern Israel. 
They also found significant anti-correlation of ionisation counts with the
solar modulation potential presenting the activity of the Sun during the
24th solar cycle. Therefore, the general conclusion of this paper could be
taken to be universally true, though the degree may be location dependent.

\section*{Acknowledgments}
The authors would like to thank the present and past balloon group team
members, such as, Dr. S. Mondal, Mr. S. Chakraborty, Mr. S.
Midya, Mr. H. Roy, Mr. R. C. Das and Mr. U. Sardar for their valuable 
supports at various stages. We also
thank Ministry of Earth Sciences (Government of India) for financial
supports to some members of the team.


\end{document}